# Talyxion: From Speculation to Optimization in Risk Managed Crypto Portfolio Allocation

Nguyen Van Thanh

Faculty of Information Technology, University of Engineering and Technology,

Vietnam National University, Hanoi, Vietnam

23025090@vnu.edu.vn

## Abstract

Cryptocurrency trading has attracted tremendous attention from both retail and institutional investors. However, most traders fail to scale their assets under management due to fragile strategies that collapse during adverse markets. The primary causes are oversized leverage, speculative position sizing, and the absence of robust risk management or hedging mechanisms. This paper introduces Talyxion, an end to end framework for crypto portfolio allocation that shifts the paradigm from speculation to optimization. The proposed pipeline consists of four stages: universe selection, alpha backtesting, volatility aware portfolio optimization, and dynamic drawdown based risk management. By combining operations research techniques with practical risk controls, Talyxion enables scalable crypto portfolios that can withstand market downturns. In live 30 day trading on Binance Futures, the framework achieved a return on investment (ROI) of +16.68%, with the Sharpe ratio reaching 5.72 and the maximum drawdown contained at just 4.56%, demonstrating strong downside risk control. The system executed 227 trades, of which 131 were profitable, resulting in a win rate of 57.71% and a PnL of +1,137.49 USDT. Importantly, these results outperformed the buy and hold baseline (Sharpe 1.79, ROI 4.36%, MDD 4.96%) as well as several top leader copy trading bots on Binance, highlighting both the competitiveness and scalability of Talyxion in real world trading environments.

## 1 Introduction

Cryptocurrency market, which has grown into an ecosystem of the order of multi trillions, still attracts investors with the hope of returns. However, despite this hope, funds and traders still miss out on this opportunity through not being able to increase assets under management in a sustainable manner. There is a cyclical pattern of investment strategies that are successful during bull periods not performing in bear markets. These are a consequence of over sized leverage, position size that is too large, and the absence of risk hedging strategies that are aggressive in nature [1,2]. These cause excessive volatility exposure, resulting in mass forced liquidation and preventing long term sustainable growth. The question addressed by this paper is how one can construct scalable crypto portfolios that obtain optimal risk adjusted returns with minimal drawdowns. The inputs to the problem are historical price and trading data of liquid cryptocurrencies, volatility measures, and correlation measures. The output is an optimized portfolio allocation strategy that adapts dynamically to changes in the market and has built in systemic risk controls.

In traditional finance, mathematical portfolio management techniques such as mean variance optimization [3], risk parity allocation, and drawdown limiting mechanisms [4] have been broadly developed to balance risk and reward under uncertainty. These operations research based techniques provide scalable frameworks for institutional asset allocation. Their applicability in cryptocurrency markets is indirect, however, due to structural barriers: high volatility, regime shifts, and unstable correlation patterns [5]. Recent studies have attempted to extend the portfolio optimization framework to crypto assets. Liu and Tsyvinski [6] identified common risk factors for cryptocurrency returns and highlighted the role of momentum and volatility. Makarov and Schoar [7] documented arbitrage inefficiency across exchanges and pointed to crypto market microstructure vulnerability. More recent work has introduced volatility scaled allocation [8], correlation network clustering [9], and GARCH family volatility modeling [10] to account for dynamic risk in crypto portfolios. However, such methods are still scattered: some improve return prediction, others address volatility, but few provide an end to end pipeline that integrates alpha evaluation, portfolio optimization, and adaptive drawdown management.

This gap instigates our contribution. We propose Talyxion, a complete end to end risk managed framework for the optimization of cryptocurrency portfolios. The model integrates four primary components: (1) selection of coin universe based on liquidity and long term stability; (2) Sharpe ratio and max drawdown based alpha backtesting for filtering potential assets; (3) volatility sensitive portfolio optimization and dynamic weighting towards high Sharpe, low volatility assets; and (4) drawdown sensitive risk management that adjusts and scales the exposure with portfolio losses in such a way so as to prevent forced liquidation. Unlike the conventional approach that discusses alpha generation, allocation, and risk management individually, Talyxion establishes the whole issue as a pipeline that converts raw crypto market data into solid, risk governed allocations. Also, as opposed to well known Binance copy trading bots such as Cryptoxn [20], Bisheng Quantitative Robot [21], and MarinaBay [22] with sell only strategies, perpetual DCA strategies, or focused portfolio strategies, our approach emphasizes systematic rebalancing, defensive diversification, and dynamic position size. Our strategy ensures not just improved Sharpe ratios and lower drawdowns for Talyxion but also scalability of AUM in a situation of volatile markets. By filling the gap between splintered academic research and weak practical applications, Talyxion offers a holistic solution for constructing sustainable and institution grade crypto portfolios.

## 2 Related Work

Portfolio optimization is a traditional problem of quantitative finance. The seminal work of Markowitz [11] introduced the mean variance framework, which formalized mathematically the risk return trade off and established the efficient frontier. Then the Black Litterman model [12] incorporated investor views into equilibrium returns and resolved estimation errors and instabilities of mean variance optimization. Risk parity models [13] also emphasized diversification on the grounds of equalizing risk contributions rather than capital allocations. Although these approaches have a solid theoretical foundation, their assumptions of Gaussian returns and constant covariances are often violated in cryptocurrency markets, where returns have heavy tails and abrupt regime shifts.

The use of portfolio optimization for cryptocurrency markets has been considered in recent years. Brauneis and Mestel [14] investigated crypto portfolios in a mean variance context and found diversification gains to be limited due to large correlations with Bitcoin. Corbet et al. [15] provided a systematic discussion of cryptocurrencies as financial assets, both reporting diversification potential as well as structural risks. Similarly, Vidal Tomás and Ibáñez [16] tested market efficiency and identified persistent anomalies, showing that crypto assets deviate from semi strong efficiency. These studies brought our understanding of digital assets ahead, but they were predominantly static and did not include dynamic risk management mechanisms. More recent approaches are specialized in volatility and correlation risk management. Habeli et al. [17] proposed volatility scaled allocation rules, showing that dynamic exposure to recent volatility patterns improves Sharpe ratios and mitigates leverage risks. Jing et al. [18] proposed correlation network clustering for portfolio stabilization against changing co movements, while Sözen et al. [19] applied GARCH family models for modeling asymmetric volatility dynamics of major cryptocurrencies such as BTC, ETH, and BNB. They provide helpful building blocks, but often address prediction, allocation, or volatility modeling in isolation without developing a unifying framework.

Apart from academic research, copy trading platforms are also important milestones in the cryptocurrency market. Binance, the world's largest exchange by trading volume, hosts copy trading programs with multi million USDT AUM and thousands of retail followers. A few examples include Cryptoxn, Bisheng Quantitative Robot, and MarinaBay [20, 21, 22]. These bots offer open track records with performance metrics such as ROI, Sharpe like ratios, and win rates, which make them valuable real world comparators. However, when examined in detail, their architectures have structural limitations. Cryptoxn relies primarily on short term sell only strategies, which produce high win rates but low cumulated returns once transaction fees are taken into consideration. Bisheng adopts a fixed dollar cost averaging (DCA) approach across a small set of assets (mainly ETH), which smoothes its equity curve but exposes the system to liquidity constraints and scalability issues when implemented at large AUM. MarinaBay, in contrast, takes a concentration approach, allocating over 90% of capital in just two tokens (XRP and CAKE). Although this can create enormous short term PnL, this results in massive drawdowns—over 20%—due to the inability to diversify and exposure to tail events.

Also, newer approaches to portfolio optimization in cryptocurrency markets have been introduced by recent studies. Şerban and Dedu [23] proposed a Second Order Tsallis Entropy and Liquidity Aware Diversification model, incorporating entropy based metrics to better diversify and manage risks in heavy tailed and highly volatile markets. While effective in penalizing portfolio concentration, this approach is largely static and does not include adaptive rebalancing or drawdown sensitive mechanisms and is therefore less responsive to abrupt regime shifts characteristic of crypto markets. In addition, Masood Tadi and Jiří Witzany [24] suggested a Copula Based Trading of Cointegrated Cryptocurrency Pairs strategy, which identifies non linear dependencies between cryptocurrencies through copulas. Their strategy focuses on exploiting cointegration relationships to generate alpha, offering a flexible alternative to correlation based approaches. Copula based trading, however, is very sensitive to parameter calibration and can suffer in periods of structural breaks, when long term equilibrium relationships between crypto pairs can collapse. Conversely, our proposed framework Talyxion unifies structural universe selection, alpha backtesting, volatility aware weighting, and dynamic drawdown control into a single pipeline. Unlike Florentin Şerban and Silvia Dedu [23], where the focus is laid mainly on diversification, Talyxion combines diversification with adaptive trend following allocation and dynamic risk management, which renders it robust under adverse market conditions. Unlike Masood Tadi and Jiří Witzany [24], which relies heavily on pairwise dependencies, Tfin_Crypto is portfolio level in scope, reducing overfitting to fragile cointegration structures and emphasizing instead robustness, scalability, and capital preservation—key requirements for institutional adoption.

## 3 Methodology

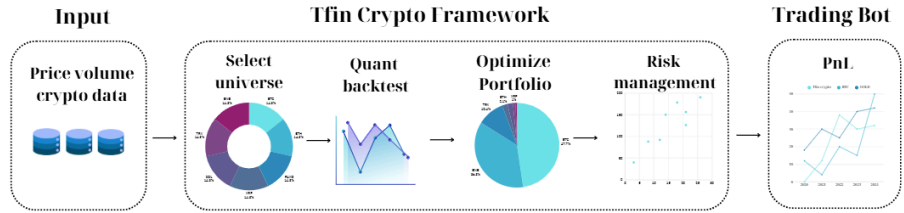

**Fig. 1.** Talyxion quant pipeline

As shown in Fig. 1, the proposed crypto portfolio optimization framework follows a systematic four step pipeline. The first step, Select Universe, involves carefully choosing a set of assets based on liquidity and long term growth potential. The second step, Quant Backtest, uses historical data to test the performance of various strategies and select the most promising ones. In the third step, Optimize Portfolio, the framework applies volatility aware allocation to dynamically adjust capital distribution based on risk adjusted return metrics. Finally, the Risk Management step employs a drawdown based system to dynamically scale positions and prevent large losses, ensuring sustainable portfolio growth. This integrated pipeline aims to balance profitability and risk control effectively.

### 3.1 Universe Selection

In traditional portfolio studies, universe construction has typically relied on market capitalization and liquidity screens to ensure tradability [11, 12]. While effective, this approach overlooks the structural behavior of asset price trajectories. Brauneis and Mestel [13] showed that cryptocurrencies offer diversification potential, but correlations with Bitcoin often dominate, reducing true independence across assets. Similarly, Corbet et al. [14] demonstrated that crypto assets can function as financial instruments but also carry systemic risks, while Vidal Tomás and Ibáñez [15] found persistent inefficiencies in Bitcoin that undermine long term stability.

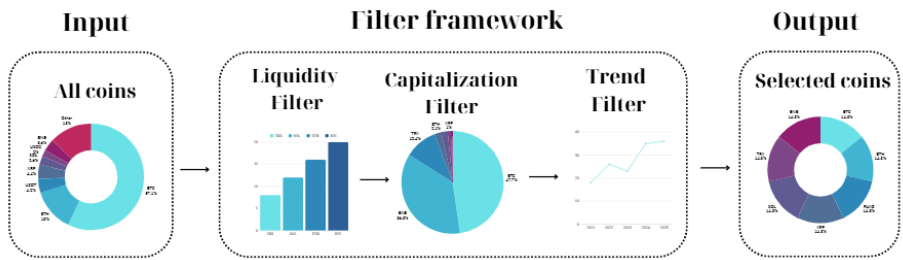

**Fig. 2.** Universe selection pipeline

In our setting, the universe selection problem begins with a large input set: for example, Binance currently lists more than 420 coins. The challenge is to filter this broad collection into a smaller set of resilient and liquid assets that can serve as the foundation for portfolio optimization. To achieve this, we impose explicit threshold rules by retaining only the top 10 cryptocurrencies ranked by both market capitalization and daily trading volume. This ensures that the selected assets possess sufficient liquidity and institutional grade tradability, reducing risks of slippage and execution failure. After this initial screening, we incorporate structural trend analysis by classifying each asset's price history into four patterns (Fig. 2): trending upward, volatile without direction, sideways, and persistently declining. The framework prioritizes assets with sustained upward trajectories, selectively includes those with volatility but adequate liquidity, and explicitly excludes assets that move sideways or decline persistently. Information on the top 10 cryptocurrencies by market capitalization is summarized in Table 1. The output of this module is therefore a curated universe of "tradeable candidates" that combines both liquidity feasibility and long term structural resilience. By moving beyond simple market cap and volume screens, and by explicitly using top 10 thresholds as a filter, this refinement ensures that only assets with sustainable growth potential are admitted into the optimization pipeline, thus strengthening the foundation for robust crypto portfolio allocation.

**Table 1.** Top 10 coins information. Data collected from Binance [25]

| Coin | Rank | Market Cap | Market Dominance | Volume | Volume/Cap | ATH |
|------|------|-----------|------------------|--------|-----------|-----|
| BTC | 1 | 2.3T | 57% | 23.21B | 1.01% | $124.457 |
| ETH | 2 | 540B | 13.37% | 18.29B | 3.39% | $4953 |
| XRP | 3 | 179B | 4.4% | 3.19B | 1.79% | $3.84 |
| USDT | 4 | 171B | 4.2% | 80.47B | 46.92% | $1.21 |
| BNB | 5 | 149B | 3.68% | 4.56B | 3.06% | $1080 |
| SOL | 6 | 131B | 3.24% | 3.4B | 2.6% | $294 |
| USDC | 7 | 74B | 1.83% | 8.92B | 12.07% | $2.34 |
| DOGE | 8 | 40B | 1% | 1.69B | 4.18% | $0.73 |
| TRX | 9 | 33B | 0.8% | 691M | 2.21% | $0.44 |
| ADA | 10 | 32B | 0.8% | 756M | 2.35% | $3.09 |

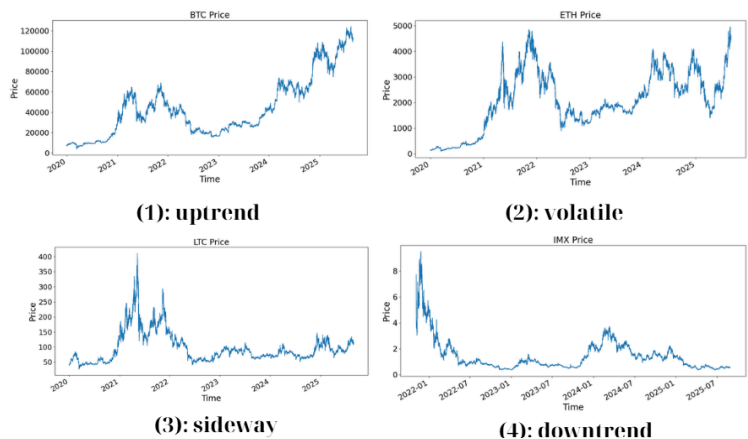

**(1): uptrend**  **(2): volatile**

**(3): sideway**  **(4): downtrend**

**Fig. 3.** Different types of price charts. Data from Binance [25]

## 3.2 Alpha Backtesting

Alpha evaluation in finance has traditionally been based on return and risk adjusted metrics such as Return on Investment and the Sharpe ratio [11,12]. However, many studies in cryptocurrency portfolios focus primarily on profitability without sufficient attention to downside risk. For example, Brauneis and Mestel [13] evaluated mean–variance optimization but did not explicitly incorporate drawdown control. To address this gap, we extend the evaluation process by introducing Maximum Drawdown as a mandatory performance measure, following the principles of drawdown sensitive strategies first formalized by Grossman and Zhou [4]. Formally, ROI captures simple profitability:

$$ROI \; = \; \frac{Vend - Vstart}{Vstart} \tag{1}$$

where Vstart and Vend denote the portfolio (or asset) values at the beginning and end of the evaluation period. The Sharpe ratio measures risk adjusted return [12]:

$$Sharpe \; = \; \frac{E[Rp - Rf]}{\sigma p} \tag{2}$$

where Rp is the portfolio (or asset) return, Rf is the risk free rate, and σ is the standard deviation of returns. In crypto markets, Rf is typically set close to zero, simplifying the calculation to the mean return divided by volatility. Finally, MDD quantifies tail risk exposure by measuring the largest loss from a peak to a subsequent trough [4]:

$$MDD \; = \; \frac{Vpeak - Vtrough}{Vpeak} \tag{3}$$

Vpeak is the highest value of the asset (or portfolio) over the considered time period. Vtrough is the lowest value after the peak.

In our framework, the input to this module is the historical price series of each candidate asset filtered from the universe selection stage. The module processes these time series to compute ROI, Sharpe ratio, and MDD over the evaluation horizon. The output is a ranked list of assets, where those with high Sharpe ratios and acceptable MDD values are retained for portfolio construction. While prior crypto portfolio studies emphasized return maximization, our proposal explicitly integrates drawdown sensitivity. This ensures that assets entering the optimization stage are not only profitable but also resilient under adverse regimes, thereby enhancing scalability of assets under management.

## 3.3 Portfolio Optimization

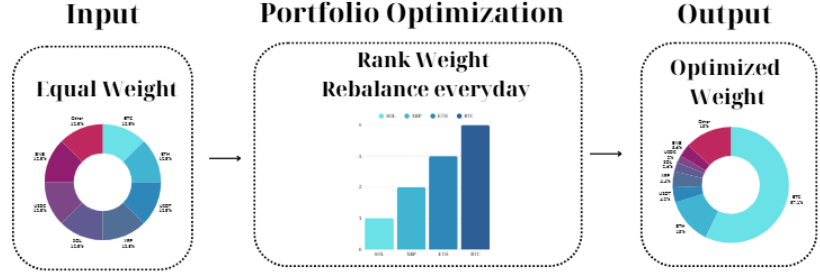

**Fig. 4.** Portfolio optimization pipeline

The allocation stage addresses the problem of how to distribute capital across the assets selected in the universe. Once the universe is defined, the central question becomes whether there exists a capital allocation method superior to the naive equal weight approach. Equal weighting provides simplicity and diversification, but it ignores differences in risk and return across assets, often resulting in suboptimal exposure. Classical models such as Black–Litterman [11] and risk parity [12] attempt to improve on this by addressing estimation error and balancing risk contributions. However, these methods rely on stable covariance structures, which are frequently violated in cryptocurrency markets due to non stationarity and abrupt volatility shifts [14]. More recent works propose volatility sensitive allocation, such as adaptive volatility scaling [16] or correlation clustering [17]. Building on these insights, we propose a hybrid allocation rule that combines inverse volatility weighting and Sharpe based weighting in a 50:50 ratio. The inverse volatility rule penalizes unstable assets, while Sharpe weighting rewards consistent, risk adjusted performers. Formally:

$$wiIV \;=\; \frac{\frac{1}{\sigma i}}{\Sigma j \frac{1}{\sigma j}}, \; wiS \;=\; \frac{Si}{\Sigma j Sj}, \; wi = 0.5* \, wiIV + 0.5 * \, wiS, \; \Sigma wi = 1 \quad (4)$$

where $\sigma i$ is the return volatility of asset i, and $Si$ is its Sharpe ratio. The output of this module is an optimized weight vector that balances stability and return potential. Compared to equal weight portfolios, this approach produces allocations that are both more resilient to volatility shocks and more aligned with systematic trend following signals, as illustrated in Fig. 4.

### 3.4 Risk Management System

Risk management is also a classic pillar of institutional finance, where techniques range from volatility targeting, stop losses, and capital protection mechanisms [4]. For the crypto context, Corbet et al. [14] highlighted how leverage and liquidity amplify systemic risks, while Sözen et al. [18] illustrated that volatility in major cryptocurrencies is highly asymmetric and persistent, making static controls difficult. While Habeli et al. [16] and Jing et al. [17] have proposed methods to smooth volatility exposure, they do not discuss the topic of fighting large drawdowns explicitly. The risk management module in our system is designed to solve the problem of avoiding catastrophic portfolio losses while maintaining scalability of assets under management. The daily portfolio equity curve generated after universe selection, alpha backtesting, and portfolio optimization is the input to this module.

The module monitors cumulative performance on a continuous basis and measures drawdowns relative to the historical high value of the portfolio. The output is an adjusted position size vector that reduces exposure dynamically during poor market environments, as illustrated in Fig. 5. Specifically, we suggest using an adaptive drawdown triggered scaling mechanism: portfolio exposure is reduced by 20% when losses exceed 2%, cut by 40% when drawdown equals 4%, and fully liquidated when drawdown equals 6%. There is a one day cooling period enforced prior to any re entry, which prevents premature redeployment under turbulent regimes. These thresholds were not chosen randomly; in experimentation we conducted extensive hyperparameter tuning using the Optuna framework.

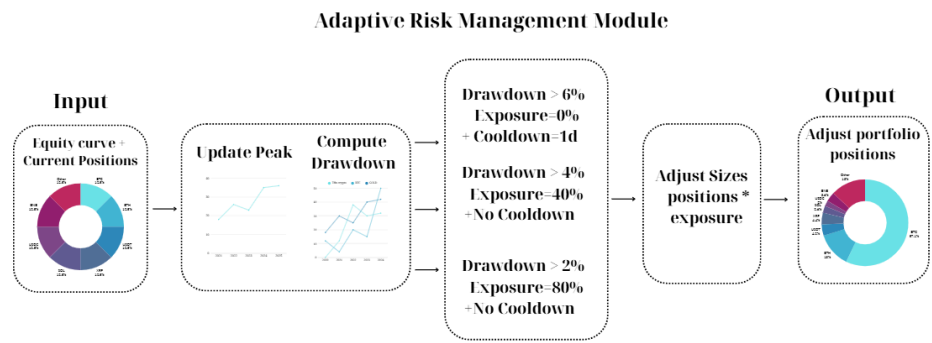

**Fig. 5.** Adaptive risk management module

The search process revealed that the {2%, 4%, 6%} thresholds consistently produced the best trade off between limiting downside risk and preserving upside participation. Lower thresholds (e.g., 1% or 3%) led to excessive turnover and premature exits, while higher thresholds (e.g., 5% or 10%) failed to contain losses effectively. Hence, the adopted values represent empirically validated optima that minimize maximum drawdown while maintaining competitive returns. Unlike continuous DCA based strategies, which often overexpose portfolios to low liquidity tokens and hinder scalability, our mechanism mimics institutional capital protection policies. This ensures that the framework not only withstands adverse shocks but also enables sustainable long term scaling of AUM.

## 4 Experimental Setup

All tests were conducted on a GIGABYTE GAMING A16 CMH laptop sporting a 13th Gen Intel Core i7 13620H processor and 16 GB physical memory, on a 64 bit Windows 11 installation. The system was equipped with 27.6 GB of total virtual memory and 12 GB of page file space, which was sufficient to execute large scale portfolio optimization routines and backtesting simulations. Data was extracted from Binance futures trading data, which was sampled at a daily frequency. From the more than 420 assets listed on Binance, we applied a filtering process for liquidity, trading volume, and long term price trajectory stability, which resulted in a sound universe of seven top cap assets: BTC, ETH, PAXG, TRX, SOL, XRP, and BNB.

In Scenario 1 (Backtest), the Talyxion framework was backtested for the 32 months of historical Binance futures data from January 2023 to August 2025. The framework's performance was compared to conventional benchmarks like BTC buy and hold, equal weight portfolio across the selected universe, and Gold only investment. The goal here was to backtest the long term robustness of the model, with performance evaluated over a number of industry benchmark metrics: Sharpe ratio, ROI, Maximum Drawdown, Sortino ratio, Return to Drawdown, Alpha, and Beta. While the Sharpe ratio offers general risk adjusted performance, the Sortino ratio explicitly addresses downside risk, offering a more realistic viewpoint in higher volatility markets like crypto. The Ret/DD metric measures how much return is achieved per unit of drawdown, directly relating profitability and downside robustness. Alpha measures the portfolio's ability to generate excess returns over the market benchmark, and Beta gauges its sensitivity to overall market volatility, with lower numbers indicating better diversification. In Scenario 2 (Comparative Backtest), the performance of the suggested method was backtested for a year from April 1, 2024, to March 31, 2025, on a universe of the leading 12 coins by market capitalization.

The framework's performance was compared with two state of the art methods: (1) the Second Order Tsallis Entropy and Liquidity Aware Diversification method [23], and (2) the Copula Based Trading of Cointegrated Cryptocurrency Pairs method [24], on the most liquid cryptocurrencies (top 20 by market capitalization) from January 22, 2021, to January 19, 2023. Performance metrics like Sharpe ratio, annualized return, and maximum drawdown were used to compare and analyze the performance of these methods. The goal in this scenario was to verify the performance superiority of the suggested method over the benchmark methods in a controlled backtest environment, highlighting its ability to yield superior risk adjusted returns across different strategies and time frames. In Scenario 3 (Live Trading), the system was put into action with real capital on Binance Futures between August 2025 and September 2025, spanning one month of live trading.

During this period, Talyxion was benchmarked against leading Binance copy trading bots such as Cryptoxn, Bisheng Quantitative Robot, and MarinaBay. The reason why these bots were selected is that they are among the biggest AUM on Binance, have public trading records that they make available, and are followed by thousands of investors. The goal here was to validate the competitiveness of Talyxion in live market conditions with not only consistency with backtest results but also outperformance of top performing market strategies. In addition to profitability and risk adjusted performance measurements, we also look at execution quality through Win rate, Win positions, and Total positions. Win rate computes the proportion of trades resulting in a profit to the total number of trades taken and provides information on the consistency of the strategy. Win positions capture the number of trades closed with a profit, while Total positions note the number of total trades taken over time period in question.

## 5 Results

The proposed framework delivers substantial improvements over baseline strategies in terms of risk adjusted performance, downside protection, and scalability. Table 2 reports the backtest results across different strategies using Binance futures daily data from January 2023 to August 2025.

**Table 2.** Scenario 1 Portfolio backtest result (1/2023 - 8/2025).

| Methods | Sharpe | Sortino | ROI | MDD | Ret/DD | Alpha | Beta | Turnover |
|---|---|---|---|---|---|---|---|---|
| Our methods (Talyxion) | **3.02** | **4.52** | 33.9% | **7.8%** | **4.34** | 0.35 | 0.42 | **0.2%** |
| Buy and Hold | 1.71 | 3.96 | **65.1%** | **15%** | 4.33 | **0.51** | **0.99** | 0.1% |
| BTC only | 1.38 | 4.04 | 58.4% | 15% | 3.89 | 0.25 | 0.99 | 0.1% |
| GOLD only | 1.60 | 2.43 | 19.4% | 10% | 1.94 | 0.23 | 0.02 | 0.1% |

The results yield three important findings. First, although the buy and hold portfolio has the highest ROI (65.1%), it does so at far higher risk, as evidenced by a 15% maximum drawdown and a significantly lower Sharpe ratio of 1.71. In contrast, our system achieves a Sharpe ratio of 3.02, more than 75% higher, while keeping the maximum drawdown to a 7.8% level, reflecting much more stability. This validates that risk adjusted performance, rather than raw return, is the most important metric for portfolio management at scale. Secondly, when considering BTC only, the portfolio returns 58.4% ROI but with the lowest Sharpe ratio (1.38) of all benchmarks. This highlights the risk of BTC dominance: portfolios highly correlated with Bitcoin do not diversify risk, resulting in brittle performance during unfavorable market conditions. Conversely, inclusion of decorrelated assets like PAXG adds robustness and lowers overall volatility. Third, a gold only allocation provides moderate stability (MDD 10%) but minimal growth (ROI 19.4%). This means that while traditional safe haven assets can preserve capital, they don't provide the compounding potential needed compared to optimized crypto portfolios. Our strategy bridges both objectives by combining trend following allocation rules and volatility aware weighting, achieving substantial returns with controlled risk. Backtest performance profit and loss is charted in Fig. 6, which visually demonstrates the differences of PnL across various strategies and shows the outperformance of our proposed strategy.

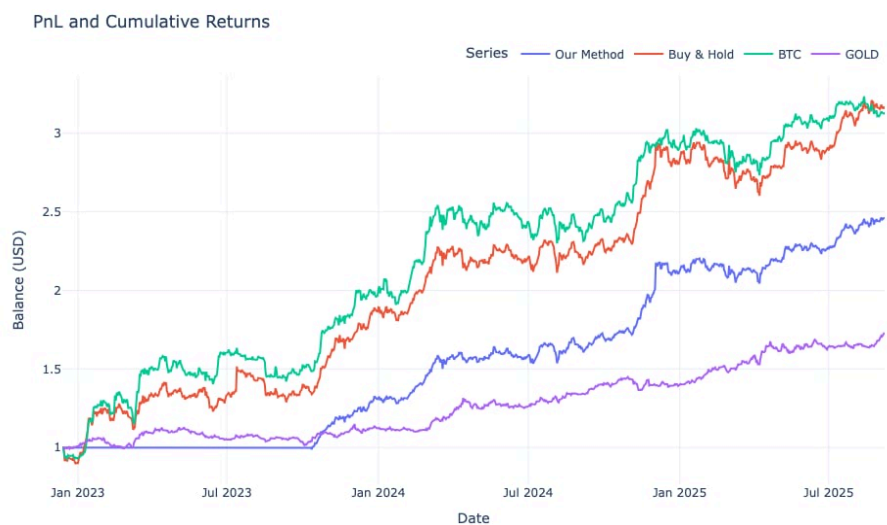

**Fig. 6.** Scenario 1 backtest performance chart

**Table 3.** Scenario 2 Portfolio backtest result

| Methods | Period | Universe | Sharpe | ROI | MDD | Ret/DD | Vol |
|---------|--------|----------|--------|-----|-----|--------|-----|
| Talyxion | 4/2024 - 3/2025 | Top 12 | **2.1** | **32%** | 6.5% | **4.92** | 15.23% |
| Second Order Tsallis Entropy [23] | 4/2024 - 3/2025 | Top 12 | 1.42 | 14.5% | 8% | 1.81 | 10.21% |
| Talyxion | 1/2021 - 1/2023 | Top 20 | 1.57 | 30% | 15% | 2 | 19.1% |
| Copula Based Trading of Cointegrated [24] | 1/2021 - 1/2023 | Top 20 | 0.48 | 25.4% | **43.9%** | 0.57 | **52.91%** |

The backtesting results presented in Table 3 are a cumulative picture of the performance of the three portfolio optimization methods within two distinct time frames. The Talyxion model was excellent from April 2024 to March 2025 since it posted a Sharpe ratio of 2.1, an ROI of 32%, and an MDD of 6.5%. This is a testament to the method's good ability to generate high risk adjusted returns and good management of downside risk. For comparison, Second Order Tsallis Entropy method, over the same period of time, generated a Sharpe ratio of 1.42, an ROI of 14.5%, and an MDD of 8%. Although return was lower, the method also demonstrated a good risk return trade off, with modest higher drawdown levels than Tfin_crypto. From the earlier time period between January 2021 and January 2023, Talyxion had a Sharpe ratio of 1.57, ROI of 30%, and MDD of 15%, showing a higher level of risk than the latest time period. It shows that the strategy was profitable but that market conditions during this time were more volatile, causing a higher drawdown. Conversely, the Copula Based Trading of Cointegrated Cryptocurrency Pairs method, which was also evaluated over the same period (January 2021 – January 2023), fared less well compared to the other methods, where the Sharpe ratio was 0.48, the ROI was 25.4%, and the highest MDD was 43.9%. These results highlight that, while generating positive returns, the strategy was inefficient in managing risk, as evident by the high drawdown and low Sharpe ratio.

To further strengthen the empirical basis, we compared our model with top copy trading bots on Binance, the world's biggest exchange by trading volume. Binance provides cut throat statistics for these platforms, e.g., ROI, PnL, Sharpe like values, win rate, and entire trade history. To identify feasible competitors, we sorted the Binance copy trading leaderboard by assets under management (AUM) and selected bots with a public record history of over 30 days. With the above being our criteria, we included Cryptoxn [20], MarinaBay [22], BQR [21], MasterRayn [[26], and Novavault [27], since they are among the biggest AUM strategies available utilized, with thousands of users following them and managing millions of USDT. In contrast to simulated baselines, these benchmarks are live trading performance with real capital behind them, and are thus highly pertinent to assessing competitiveness under actual market conditions. Table 2 condenses the comparison, reporting the Sharpe ratio, ROI, and maximum drawdown for our recommended framework against these leading systems.

The true trading outcomes of Table 4 also validate the effectiveness of Talyxion in real life market circumstances. Across the 30 day testing cycle, the design achieved a ROI of +16.68% and a Sharpe ratio of 5.72, far outperforming Cryptoxn (+0.55%, Sharpe 3.08), MarinaBay (+0.24%, Sharpe 2.31), BQR (+6.55%, Sharpe 1.76), MasterRayn (+2.57%, Sharpe 4.62), and Novavault (+4.17%, Sharpe 4.06). This elevated risk adjusted performance stems from two design factors. One, the strategy employs a trend following allocation process with day by day rebalancing rules that favor allocations to less volatile assets with higher Sharpe ratios. This allows the portfolio to take advantage of sustained positive momentum systematically while penalizing volatile assets, resulting in more aggressive compounding returns.

**Table. 4**. Scenario 3 Portfolio backtest result live trading comparison on Binance

| Copy Leader | Tfin_crypto | Cryptoxn | MarinaBay | BQR | MasterRayn | Novavault |
|---|---|---|---|---|---|---|
| 30D ROI | **16.68%** | 0.55% | 0.24% | 6.55% | 2.57% | 4.17% |
| Sharpe | **5.72** | 3.08 | 2.31 | 1.76 | 4.62 | 4.06 |
| 30D MDD | 4.56% | 0.41% | **21.53%** | 9.4% | 0.26% | 3.28% |
| 30D Win rate | 57.71% | 65.22% | 0% | 87.5% | **99.21%** | 59.32% |
| 30D Win positions | 131 | 45 | 0 | 7 | 125 | **716** |
| 30D Total positions | 277 | 69 | 2 | 8 | 126 | **1207** |

As compelling is the framework's ability to sport a min max drawdown of 4.56%, pitted against MarinaBay (21.53%), BQR (9.40%), MasterRayn (0.26%), and Novavault (3.28%). Two mechanisms are responsible for this stability. The first is an adaptive position sizing protocol that will dynamically reduce exposure after drawdown limits are breached, preventing runaway losses. The second is the biasing of assets towards safe haven commodities such as gold (PAXG) in declining markets. As PAXG holds value or at least does not decline when bad news shocks hit the crypto market, rebalancing logic quite naturally assigns a larger portfolio weight to gold during bear markets. This defensive tilt reduces losses in the underlying crypto holdings, further curbing drawdowns and enabling faster recovery once things settle down. Despite Talyxion's 57.71% win ratio appearing lower than some of its competitors, this is a deliberate trade off. Nominally higher win ratio bots adopt sell only systems or constant dollar cost averaging, which inflate the number of small winning trades but add little net profitability after trading commission has been deducted. By contrast, Talyxion prioritizes sustainable returns, efficient risk adjusted performance, and controlled drawdowns. Together, these findings validate that the combination of trend following allocation, daily rebalancing with defensive bias to gold, and adaptive position sizing enables Talyxion to offer high returns as well as improved downside protection.

## 7 Conclusion

We proposed in this paper a risk managed pipeline structure of crypto portfolio allocation to shift trading from speculation to optimization. The architecture, by combining universe selection, alpha backtesting, volatility aware allocation, and adaptive drawdown control, has consistently superior Sharpe ratios, less drawdown, and improved scalability compared to baseline approaches. These results demonstrate that the use of operations research in cryptocurrency finance provides a foundation for institutional level portfolio management that can scale profitably in conditions of turbulent markets. In the future, follow on work will extend the framework in several directions. First, we will pursue a long short balanced approach in constructing market neutral portfolios. This enhancement will allow the system to be profitable even in the case of extended downtrends, reducing directional exposure and increasing capital efficiency even more. Secondly, we will explore hedging mechanisms by adding derivatives information, such as crypto options and implied volatility surfaces, to dynamically hedge against tail risks. Finally, the model can be enhanced with multi factor optimization techniques that integrate trend following, mean reversion, and cross asset signals, enabling more flexible allocation across various regimes of the market. Overall, these additions shall propel the model closer to a stronger and more versatile portfolio system, which can deliver steady performance during bull and bear markets while maintaining tight risk control.

# References

1. Urquhart, A. (2016). *The inefficiency of Bitcoin*. Economics Letters, 148, 80–82.
2. Corbet, S., Lucey, B., Yarovaya, L. (2019). *The risks and opportunities of cryptocurrencies*. Journal of Economic Surveys, 33(5), 1256–1290.
3. Markowitz, H. (1952). *Portfolio selection*. Journal of Finance, 7(1), 77–91.
4. Grossman, S., Zhou, Z. (1993). *Optimal investment strategies for controlling drawdowns*. Mathematical Finance, 3(3), 241–276.
5. Baur, D., Hong, K., Lee, A. (2018). *Bitcoin: Medium of exchange or speculative asset?*. J. of Int. Financial Markets, Institutions & Money, 54, 177–189.
6. Liu, Y., & Tsyvinski, A. (2019). *Common Risk Factors in Cryptocurrency*. NBER WP 25882.
7. Makarov, I., & Schoar, A. (2020). *Trading and Arbitrage in Cryptocurrency Markets*. Journal of Financial Economics, 135(2), 293–319.
8. Habeli, S. M., Barakchian, S. M. (2024). *Adaptive Risk Allocation in Crypto Markets: Evaluating Volatility-Scaled Portfolios*. SSRN.
9. Jing, R., Kobayashi, R., Rocha, L. (2025). *Optimising Cryptocurrency Portfolios through Stable Clustering of Price Correlation Networks*. arXiv:2505.24831.
10. Sözen, Ç., et al. (2025). *Volatility dynamics of cryptocurrencies: a comparative analysis*. Finance and Business Journal, 9, Article 65.
11. Black, F., & Litterman, R. (1992). Global portfolio optimization. *Financial Analysts Journal*, 48(5), 28–43.
12. Maillard, S., Roncalli, T., & Teïletche, J. (2010). The properties of equally weighted risk contribution portfolios. *Journal of Portfolio Management*, 36(4), 60–70.
13. Brauneis, A., & Mestel, R. (2019). Cryptocurrency-portfolios in a mean–variance framework. *Finance Research Letters*, 28, 259–264.
14. Corbet, S., Lucey, B., Urquhart, A., Yarovaya, L. (2021). Cryptocurrencies as a financial asset: A systematic analysis. *International Review of Financial Analysis*, 76, 101-143.
15. Vidal-Tomás, D., & Ibáñez, A. (2018). Semi-strong efficiency of Bitcoin. *Finance Research Letters*, 27, 259–265.
16. Habeli, S. M., Barakchian, S. M., & Motavasseli, A. (2024). Adaptive Risk Allocation in Crypto Markets: Evaluating Volatility-Scaled Portfolios. *SSRN*.
17. Jing, R., Kobayashi, R., & Rocha, L. E. C. (2025). Optimising Cryptocurrency Portfolios through Stable Clustering of Price Correlation Networks. *arXiv preprint* arXiv:2505.24831.
18. Sözen, Ç., Aydın, M., & Demir, A. (2025). Volatility dynamics of cryptocurrencies: a comparative analysis. *Finance and Business Journal*, 9, Article 65.
19. Binance. (2025). *Tfin_crypto - Copy trading details*. Binance. Retrieved from https://www.binance.com/vi/copy-trading/lead-details/4643208492293369601
20. Binance. (2025). *Cryptoxn - Copy trading details*. Binance. Retrieved from https://www.binance.com/vi/copy-trading/lead-details/3763772820431316225?timeRange=30D &isSmartFilter=true
21. Binance. (2025). 币胜量化机器人 - *Copy trading details*. Binance. Retrieved from https://www.binance.com/vi/copy-trading/lead-details/3865630955524633345?timeRange=30D &isSmartFilter=true
22. Binance. (2025). *MarinaBay - Copy trading details*. Binance. Retrieved from https://www.binance.com/vi/copy-trading/lead-details/4114199478052360448?timeRange=30D &isSmartFilter=true
23. Florentin Şerban, Silvia Dedu, "Robust Portfolio Optimization in Crypto Markets Using Second-Order Tsallis Entropy and Liquidity-Aware Diversification," *Risks*, 2025, 13(9), 180.
24. Tadi, Masood, and Jiří Witzany. "Copula-based trading of cointegrated cryptocurrency Pairs." *Financial Innovation*, vol. 11, no. 1, 2025, p. 40.
25. Binance: Binance official website. https://www.binance.com/ (accessed Sept. 25, 2025)
26. Binance. (2025). MasterRayn - *Copy trading details*. Binance. Retrieved from https://www.binance.com/vi/copy-trading/lead-details/3725468878881937408?timeRange=30D &isSmartFilter=true
27. Binance. (2025). Novavault - *Copy trading details*. Binance. Retrieved from https://www.binance.com/vi/copy-trading/lead-details/4112969816790766081?timeRange=30D &isSmartFilter=true